\documentclass[onecolumn,amsmath,amssymb,prd,reprint,longbibliography,floatfix,8pt,a4paper,aps]{revtex4-1}

\usepackage{cancel}
\usepackage{enumitem}
\usepackage{mathtools}
\usepackage[utf8x]{inputenc}
\usepackage{graphicx}
\usepackage{dcolumn}
\usepackage{bm}
\usepackage{placeins}
\usepackage[switch]{lineno}
\usepackage{float} 
\usepackage{subfigure}
\usepackage{tikz}
\setcitestyle{super}
\usepackage{multirow}
\usetikzlibrary{arrows.meta}
\usetikzlibrary{arrows,decorations.pathmorphing,backgrounds,positioning,fit,petri}

\begin{document}

\title{Spin decoherence in molecular crystals: nuclear v.s. electronic spin baths}

\author{Conor Ryan}
\author{Valerio Briganti}
\author{Cathal Hogan}
\author{Mark O'Neill}
\author{Alessandro Lunghi}
\email{lunghia@tcd.ie}
\affiliation{School of Physics, AMBER and CRANN Institute, Trinity College, Dublin 2, Ireland}

\begin{abstract}
\noindent
The loss of information about the relative phase between two quantum states, known as decoherence, strongly limits resolution in electron paramagnetic spectroscopy and hampers the use of molecules for quantum information processing. At low temperatures, the decoherence of an electronic molecular spin can be driven by its interaction with either other electron spins or nuclear spins. Many experimental techniques have been used to prolong the coherence time of molecular qubits, but these efforts have been hampered by the uncertainty about which of the two mechanisms is effectively limiting coherence in different experimental conditions. Here, we use the Cluster-Correlation Expansion (CCE) to simulate the decoherence of two prototypical molecular qubits and quantitatively demonstrate that nuclear spins become the leading source of decoherence only when the electron spin concentration is below $\sim$ 1 mM. Moreover, we show that deuterated samples, much easier to achieve than fully spin-free environments, could achieve record coherence times of $\sim$ 0.1 ms for an electron spin concentration of $\sim$ 0.1 mM. Alternatively, hydrogen-rich molecular crystals with electron spins concentration below 1 mM can still achieve coherence times of 10 ms through dynamical decoupling, showing that the potential of molecular spins for quantum technologies is still untapped. 
\end{abstract}

\maketitle

\section*{Introduction}
\noindent
Magnetic molecules with a doublet ground state ($S=1/2$), either originating from organic radicals or coordination compounds, have long been used as probes of their unpaired electron's environment and more recently have surged as promising building blocks for quantum technology\cite{moreno2018molecular,coronado2020molecular,wasielewski2020exploiting,lavroff2021recent}. This promise is highlighted through properties such as the ability to control the qubits' properties by tuning the chemical nature of its magnetic environment and the potential to synthesize many identical molecular qubits and deposit them in ordered arrays\cite{gaita2019molecular,fursina2023toward}.

An integral feature of spin qubits is their coherence time, denoted by $T_2$, which governs how long the molecular spin will retain its quantum behavior and, therefore, can have quantum operations performed on it during, for instance, computation or signal detection in sensing. Often, $T_2$ is decomposed in terms of two different contributions: the spin relaxation time, $T_1$, and pure dephasing time, $T_2^*$. According to perturbation theory arguments\cite{yang2016quantum,mondal2023spin}, all these quantities are related as
\begin{equation}
    \frac{1}{T_2}=\frac{1}{2T_1}+\frac{1}{T^*_2}\:.
\end{equation}
Generally speaking, $T_1$ is the time the spin needs to reach its equilibrium state by exchanging energy with a surrounding bath, while $T_2^*$ describes the loss of phase coherence among identical spins in an ensemble due to perceiving a different dynamically changing environment. Static inhomogeneous contributions to $T_2$ are here neglected as they can be easily removed with a simple Hahn-Echo pulse sequence\cite{hahn1950spin,yang2016quantum} and do not pose particular challenges.

At high temperatures, often already above 10-20 K, spin-phonon interactions are the leading cause of decoherence in solid-state spin systems, be they color centers like NV centers\cite{bar2013solid,mondal2023spin} and vacancies in h-BN\cite{gottscholl2021spin,liu2022spin}, impurities such as P-doped Si\cite{melnikov2004quantum,jeong2010spin}, or molecular qubits\cite{bader2014room,zadrozny2015millisecond,atzori2016room}. At the microscopic level, this interaction arises from the modulation of the electronic structure defining the qubit by lattice vibrations and is made possible by spin-orbit coupling. Spin-phonon coupling has been the focus of intense research in recent years\cite{lunghi2019phonons,garlatti2023critical,espinosa2025slow,eaton2025anisotropy}, and its contribution to both $T_1$ and $T_2^*$ is now largely established\cite{mondal2023spin,lunghi2023spin,mariano2025role}.

On the other hand, at low temperatures, where the phonon population eventually vanishes, spin-phonon decoherence is suppressed, and $T_2$ becomes dominated by spin-spin dynamics\cite{takahashi2011decoherence}. The latter originates from the interaction of the qubit's spin moment with other electronic spins, often other equivalent spin qubits present in the surrounding environment, or nuclear spins. The latter interaction is generally considered as prominent in molecular qubits, where hydrogen atoms, with their large nuclear spin gyromagnetic factor, are commonly present in the qubit's structure itself and its solid-state matrix.

Several successful strategies have been employed to date to counteract spin-spin decoherence. For instance, dynamical decoupling has been used in a few instances to enhance $T_2$ by systematically refocusing spins precessing at different rates and thus subject to pure dephasing\cite{soetbeer2018dynamical,dai2021experimental,pazera2023pulse}. The engineering of spin state properties, such as introducing clock transitions\cite{shiddiq2016enhancing, tlemsani2025assessing}, is another possible strategy employed in recent years to protect coherence from magnetic noise. Finally, chemical principles have also been employed in the attempt to dilute the spin bath as much as possible. This is relatively easy to perform for the electronic spin bath by diluting qubits in a solid-state diamagnetic host or frozen solutions. Removing nuclear spins is also possible\cite{bader2014room,zadrozny2015millisecond}, but at the cost of drastically reducing the type of organic ligands that can be possibly used to assemble magnetic molecules. As a compromise, the substitution of hydrogen for deuterium, which bears a much smaller gyromagnetic factor, is a commonly employed strategy\cite{soetbeer2018dynamical}.

Despite the abundant evidence that dilution strategies work, a systematic study on the role of both electronic and nuclear spin baths is currently lacking, and incongruous reports on the value of $T_2$ in different spin environments are present in the literature. For instance, the works of Bader et al.\cite{bader2014room} and Zadrozny et al.\cite{zadrozny2015millisecond} both report molecular qubits based on hydrogen-free coordination compounds, but found starkly different values of $T_2$ for deuterated molecular environments, 68 $\mu$s and 8 $\mu$s, respectively. Similarly, the work of de Camargo et al.\cite{de2021exploring} reported a molecular qubit in deuterated frozen solvent to have a $T_2$ of about 10-40 $\mu$s, which is surprisingly longer than what was observed by Zadrozny et al.\cite{zadrozny2015millisecond} for a hydrogen-free molecular qubit. To complicate the matter, all these experiments are performed at different molecular spin concentrations, which may vary from $\sim$ 45 mM\cite{chicco2021controlled} to 0.01 mM\cite{zadrozny2015millisecond}. On this last regard, while experimental studies often comment on the innocence of electronic spin baths in their respective operating conditions\cite{zadrozny2015millisecond}, systematic evidence for such claims is often not available, and a theoretical study suggested the opposite\cite{lunghi2019electronic}, with interactions between the qubit and its electronic spin bath predicted to be driving spin decoherence for concentrations down to $\sim$ 0.5 mM.

This state of affairs obscures the true nature of the mechanism that leads spin decoherence at low temperatures, severely impacting our ability to judge the real value of different strategies for improving $T_2$, and ultimately precluding us from maximizing spin coherence.

Theory and computational modeling can play a key role in solving these ambiguities. Here we build on the corpus of literature employing Cluster-Correlation Expansion (CCE) methods\cite{yang2008quantum} and similar approaches to tackle the problem of simulating spin decoherence\cite{ma2014uncovering,canarie2020quantitative, chen2020decoherence,ghosh2021spin, onizhuk2021probing,jahn2024contribution}. Unlike previous reports\cite{lenz2017quantitative, canarie2020quantitative, chen2020decoherence}, here we simulate on the same footing the effect of both electron and nuclear spin baths over the coherence of molecular spins. We perform our study for two prototypical molecular qubits with long coherence times in molecular crystals: VO(TPP)\cite{yamabayashi2018scaling} and (PPh$_4$)$_2$[Cu(mnt)$_2$]\cite{bader2014room, lenz2017quantitative}. These molecular crystals offer an ideal playground for our investigation for several reasons. To start with, they have been investigated in diluted crystals, making it possible to know the position of hydrogen atoms with atomic precision. Moreover, they exhibit quite different chemical environments, with only molecular hydrogen atoms being present in VO(TPP), and only counter-ion hydrogen atoms being present for the hydrogen-free molecule [Cu(mnt)$_2$]$^{2-}$. In terms of experimental conditions, the two studies present significant variations. For instance, $T_2$ has been measured for VO(TPP) at the high concentrations of 44.65 mM\cite{yamabayashi2018scaling}, while [Cu(mnt)$_2$]$^{2-}$ has been further diluted down to 0.15 mM and 0.015 mM\cite{bader2014room}. Finally, the (PPh$_4$)$_2$[Cu(mnt)$_2$] crystal has also been studied in deuterated samples\cite{bader2014room}, making it possible to fully address the broad phenomenology of spin baths.

Our study demonstrates the critical importance of electronic spin baths in the decoherence of molecular spins. Importantly, this new knowledge not only makes it possible to determine the crossover concentration at which electron and nuclear spin baths swap in relevance and reinterpret several experimental studies, but also establishes that fully deuterated samples can achieve outstanding $T_2$ values of $\sim$ 0.1 ms if their electronic spin baths are properly diluted, which are very close to the coherence time of 0.7 ms obtained for spin-free samples\cite{zadrozny2015millisecond}. This upper limit for deuterated compounds has so far remained elusive to experimental studies that have invariably employed electronic spin baths at too high concentrations and demonstrate that the potential of molecular qubits for quantum information processing is still untapped. Finally, we simulate dynamical decoupling experiments for both electronic and nuclear baths, establishing the ultimate limits of coherence times in $S=1/2$ molecular crystals. 

\section*{Theory}
\noindent
In this section, the theory behind the contributions to the decoherence of a molecular spin qubit through spin-spin interactions is discussed, along with an explanation of the cluster-correlation expansion method as a way to tractably simulate spin decoherence due to a spin bath open quantum system.

\subsection*{Spin Decoherence}
\noindent
A central spin $\vec{\textbf{S}}$, interacting with a spin bath $\{\vec{\textbf{I}}_i\}$, in a magnetic field $\vec{\textbf{B}}$, is described by the following spin Hamiltonian\cite{onizhuk2021pycce},
\begin{align}
    \hat{H} &= \vec{\textbf{S}}\cdot\textbf{D}\cdot\vec{\textbf{S}}+\vec{\textbf{B}}\cdot\gamma_S\cdot\vec{\textbf{S}}+\sum_i\vec{\textbf{S}}\cdot\textbf{A}_i\cdot\vec{\textbf{I}}_i+\sum_i\vec{\textbf{I}}_i\cdot\vec{\textbf{P}}_i\cdot\vec{\textbf{I}}_i\notag\\&+\vec{\textbf{B}}\cdot\gamma_i\cdot\vec{\textbf{I}}_i+\sum_{i<j}\vec{\textbf{I}}_i\cdot\textbf{J}_{ij}\cdot\vec{\textbf{I}}_j,
\end{align}
with $\textbf{D}(\textbf{P)}$ being the zero-field splitting (quadrupole) tensor of the central (bath) spin, $\textbf{A}$ the interaction tensor between the central and bath spins, $\textbf{J}$ the interaction tensor between bath spins, and $\gamma$ the interaction tensor between the spin and the magnetic field. In general, time evolution under this Hamiltonian will induce both relaxation and pure dephasing contributions to the decoherence of the central spin, highlighted by the decomposition of $T_2$ into contributions from $T_1$ and $T_2^*$ in Eq (1). First, considering the relaxation contribution, the existence of non-zero Hamiltonian matrix elements such as $\langle+|\hat{H}|-\rangle$ would result in non-zero transition amplitudes between the spin states of the central spin, $\{|+\rangle, |-\rangle\}$. Importantly, this case of non-zero transition elements of the Hamiltonian becomes operative when the splitting of the central spin energy levels is similar to that of the bath spins, therefore enhancing the probability of a spin flip from the central spin transferring spin polarization to the bath. This is particularly likely to occur in situations where the spin species of both the central spin and the bath spins are the same, for example, an electron central spin in a bath of electron spins, since they will all be similarly affected by the magnetic field. The timescale at which this transfer of spin polarization occurs is known as $T_1$, the spin-relaxation time.

Next, considering either the dephasing contribution to decoherence in a system that experiences both relaxation and dephasing (electron central spin in a bath of electron spins), or decoherence in a system with $T_1\gg T_2$ that only experiences pure dephasing (electron central spin in a bath of nuclear spins), the Hamiltonian can be projected onto the central spin degrees of freedom,
\begin{equation}
    \hat{H} = |+\rangle\langle+|\otimes\hat{H}^{(+)}+|-\rangle\langle-|\otimes\hat{H}^{(-)},
\end{equation}
using the assumption that $\langle+|\hat{H}|-\rangle=\langle-|\hat{H}|+\rangle=0$, i.e. neglecting relaxation effects just for the purpose of isolating the dephasing behavior and studying its individual effect on decoherence. In fully general cases neither of

Performing this projection on the spin Hamiltonian, the operators $\hat{H}^{(\pm)}$ are obtained which are bath spin Hamiltonians conditioned on the state of the central spin. This then has the following effect on the time evolution operator,
\begin{equation}
    e^{-i\hat{H}t} = |+\rangle\langle+|\otimes e^{-i\hat{H}^{(+)}t} + |-\rangle\langle-|\otimes e^{-i\hat{H}^{(-)}t},
\end{equation}
meaning the dynamics of the bath now becomes conditional on the state of the central spin.

Considering the initial central spin state to be $|\psi\rangle=(|+\rangle+|-\rangle)/\sqrt{2}$, as is standard when studying spin decoherence, and taking the bath to be in some pure state $|\mathcal{J}\rangle$, using Eq (4), the combined spin-bath state evolves as
\begin{equation}
    \frac{|+\rangle+|-\rangle}{\sqrt{2}}\otimes|\mathcal{J}\rangle\xrightarrow[]{e^{-i\hat{H}t}}\frac{|+\rangle\otimes|\mathcal{J}_+(t)\rangle}{\sqrt{2}}+\frac{|-\rangle\otimes|\mathcal{J}_-(t)\rangle}{\sqrt{2}},\notag
\end{equation}
where the resulting state now contains two separate bath states, $|\mathcal{J}_{\pm}(t)\rangle$ = $e^{-i\hat{H}^{(\pm)}t}|\mathcal{J}\rangle$, that are conditional on the state of the central spin, meaning that this is clearly an entangled state between the central spin and the bath.

Expressed through these results, the coherence of the central spin can be written as the overlap between the differently evolved bath states,
\begin{equation}
    \mathcal{L}(t)=\langle\mathcal{J}_-(t)|\mathcal{J}_+(t)\rangle=\langle\mathcal{J}|e^{i\hat{H}^{(-)}t}e^{-i\hat{H}^{(+)}t}|\mathcal{J}\rangle.
\end{equation}
Information on the state of the central spin is then encoded in the evolution of the bath states, and as the evolution of these states is governed by different Hamiltonians there will be a decay in their overlap and hence the coherence. This decay is characterized by the pure dephasing time, $T_2^*$. Generally, the initial coherence of the central spin will then evidently decay as a result of both relaxation and dephasing contributions induced by spin-spin interactions with the spin bath. 

\subsection*{Cluster-Correlation Expansion (CCE)}
\noindent
Explicitly considering time evolution governed by the Hamiltonian given in Eq (2), the time evolution operator can be expanded as a Taylor series in the spin-bath coupling strength. From here, the arguments of perturbation theory can be applied under the assumption that higher order terms in the expansion are weaker in powers of the spin-bath coupling strength. There has previously been a focus on calculating the central spin decoherence through evaluating each term of the perturbative expansion using a Feynman diagram style approach known as the Linked-Cluster Expansion(LCE)\cite{saikin2007single}. While LCE proved to be successful in calculating spin decoherence, as more diagrams become necessary at higher orders, as well as for spins greater than 1/2, the process rapidly becomes tedious and more efficient methods become sought after.

At each order in the spin-bath coupling strength, say order $k$, the corresponding term in the perturbative expansion contains spin-spin interactions of clusters of bath spins containing up to $k$ spins. This knowledge, combined with the intuition that the timescale at which decoherence occurs is typically much shorter than the time it takes for spin flip-flops, processes in which two spins of any type exchange their quantum states, resulting in them flipping their spin orientations without any change in their total energy, to build up correlations among a large number of spins in the bath, means that the perturbative expansion of the time evolution operator can be rearranged into terms ordered by the size of the bath spin clusters rather than the order of the system-bath coupling strength. The coherence of the central spin can then be decomposed into contributions from correlated clusters of bath spins of increasing size. This decomposition is known as the Cluster-Correlation Expansion\cite{yang2008quantum}. The approach of CCE is to partition the spin bath into overlapping clusters of increasing size up to some truncation size $M$, where generally $M\ll N$ for a bath of $N$ spins. This allows the coherence of the central spin to be accurately approximated as a product of irreducible cluster coherence functions up to the truncation size,
\begin{equation}
    \mathcal{L}^{(M)}(t) = \prod_{|\mathcal{C}|\leq M}\tilde{L}_\mathcal{C}(t).
\end{equation}
The irreducible coherence functions can be calculated recursively from the coherence functions from each bath cluster,
\begin{equation}
    \tilde{L}_\mathcal{C}(t) = \frac{L_\mathcal{C}(t)}{\prod_{\mathcal{C'}\subset\mathcal{C}}\tilde{L}_\mathcal{C'}(t)},
\end{equation}
\begin{equation}
    L_\mathcal{C}(t) = \langle+|e^{-i\hat{H}_\mathcal{C}t}\rho_{S+\mathcal{C}}(0)e^{i\hat{H}_\mathcal{C}t}|-\rangle.
\end{equation}
The irreducible coherence functions, $\tilde{L}_\mathcal{C}(t)$, are then measures of the correlations of the bath spins arising only due to spin flip-flops of all bath spins in the clusters $\mathcal{C}$, since the correlations of their subclusters are factored out.

The cluster Hamiltonian which governs the time evolution of the central spin-cluster combined system is given by,
\begin{align}
    \hat{H}_\mathcal{C} &= \vec{\textbf{S}}\cdot\textbf{D}\cdot\vec{\textbf{S}}+\vec{\textbf{B}}\cdot\gamma_S\cdot\vec{\textbf{S}}+\sum_{i\in\mathcal{C}}\vec{\textbf{I}}_i\cdot\textbf{P}_i\cdot\vec{\textbf{I}}_i+\vec{\textbf{B}}\cdot\gamma_i\cdot\vec{\textbf{I}}_i\notag\\&+\sum_{i<j\in\mathcal{C}}\vec{\textbf{I}}_i\cdot\textbf{J}_{ij}\cdot\vec{\textbf{I}}_j+\sum_{i\in\mathcal{C}}\vec{\textbf{S}}\cdot\textbf{A}_i\cdot\vec{\textbf{I}}_i+\sum_{a\notin\mathcal{C}}\vec{\textbf{S}}\cdot\textbf{A}_a\langle\vec{\textbf{I}}_a\rangle\notag\\&+\sum_{i\in\mathcal{C},a\notin\mathcal{C}}\vec{\textbf{I}}_i\cdot\textbf{J}_{ia}\langle\vec{\textbf{I}}_a\rangle.
\end{align}
This Hamiltonian consists of all terms for the central and cluster bath spins, with the non-cluster bath spins included in the last two terms as a mean-field average. These mean-field average terms then allow for the effect of the non-cluster bath spins to be included in a straightforward manner as a static field contribution to the dynamics. The explicit inclusion of the central spin degrees of freedom in the Hamiltonian means this approach in particular is known as the generalized Cluster-Correlation Expansion (gCCE)\cite{onizhuk2021pycce}, and is capable of simulating relaxation effects on the dynamics of the decoherence. Due to its full generality, gCCE will be used to produce all computational result in this work. The cluster bath spins are included dynamically through their spin flip-flops causing fluctuations of the local-field experienced by the central spin, resulting in decoherence. All spin-spin interactions are also approximated as interactions between point dipoles.

Taking each bath spin to interact on average with $q$ other bath spins, with a typical pair flip-flop interaction strength of $\alpha_I$, a truncated gCCE will converge at times $T$ which satisfy $q\alpha_IT\ll1$\cite{yang2008quantum}. However, convergence at timescales beyond this range may still be possible.

In summary, CCE decomposes the dynamics of the central spin decoherence into contributions from overlapping correlated clusters of bath spins which arise as a result of spin flip-flops involving all bath spins in the cluster, thus dynamically altering the local magnetic field of the central spin. The problem is then reduced from simulating the dynamics of the very large Hilbert space of the full bath, to one requiring the much more computationally tractable repeated simulation of reduced bath cluster Hilbert spaces.

\section*{Results}
\noindent
Using the PyCCE Python package\cite{onizhuk2021pycce}, the individual contributions to the spin decoherence of the VO(TPP) and [Cu(mnt)$_2$]$^{2-}$ transition metal complex molecular qubits is simulated using gCCE. A quantitative analysis is then performed to calculate the optimal target concentration of the electronic spin bath which will suppress its contribution to decoherence in favor of the contributions of the nuclear spins in the bath, which will result in longer values of the coherence times of the qubits.

\subsection*{Decoherence in VO(TPP)}
\noindent
The VO(TPP) (TPP = tetraphenylporphyrinate) molecular spin qubit contains a single unpaired spin-1/2 electron in the d-shell of the V ion. The crystal unit cell contains two molecular units as shown in Figure 1. Given the neutral electric charge of the molecule, no counter-ions are present in the system. The crystal structure used in this work is an optimized structure found through DFT calculations\cite{garlatti2023critical}.
\begin{figure}[htp]
    \centering
    \includegraphics[width=8cm]{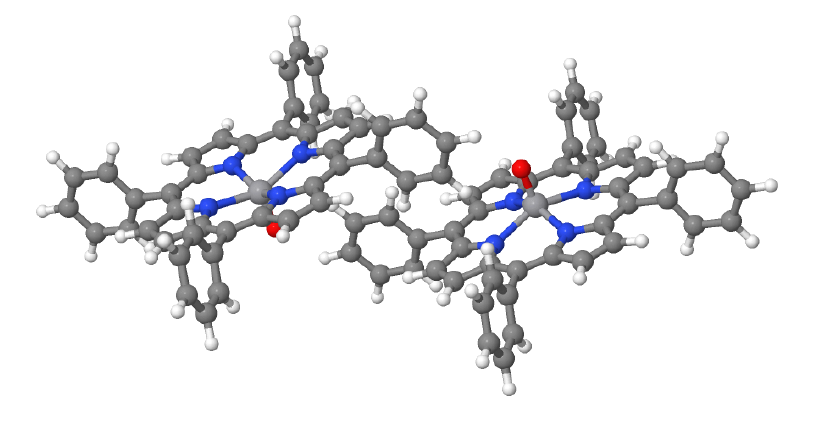}
        \label{fig:VOTPP}
    \caption{\textbf{ VO(TPP) crystal structure.} The atoms in the crystal unit cell are reported with the color code: vanadium (light grey), oxygen (red), nitrogen (blue), carbon (dark grey), hydrogen (white).}
\end{figure}
\FloatBarrier
\noindent
The electron spin interacts with the spin-7/2 $^{51}V$ nuclear spin and the external magnetic field through the Hamiltonian\cite{chicco2021controlled}
\begin{equation}
    \hat{H} = \mu_B\vec{\textbf{B}}\cdot\textbf{g}_S\cdot\vec{\textbf{S}} + \vec{\textbf{S}}\cdot\textbf{A}\cdot\vec{\textbf{I}} + \mu_Ng_N\vec{\textbf{B}}\cdot\vec{\textbf{I}} + pI_z^2 \:,
\end{equation}
with the following hyperfine coupling, \textbf{A}, and electronic $g$-tensors previously obtained from DFT calculations\cite{garlatti2023critical},
\begin{table}[htp]
\begin{center}
    \label{table:VOTPP_H}
    \caption{Hamiltonian Parameters for VO(TPP)}
    \begin{tabular}{ |p{2cm}|p{2cm}|p{2cm}|p{2cm}|  }
     \hline
     $g_{\perp}$ & $g_{\parallel}$ & $A_{\perp}$ (MHz) & $A_{\parallel}$ (MHz)\\
     \hline
     1.984 & 1.968 & -166 & -473\\
     \hline
    \end{tabular}
\end{center}
\end{table}
\noindent
and the quadrupolar coupling of the $^{51}$V nucleus, $p$ = $-$0.35 MHz, measured from its spectrum as a function of the applied magnetic field\cite{chicco2021controlled}. The NMR parameters for the $^{51}V$ nucleus, $g_N$ and $\mu_N$, are 1.46837 and 7.6227 MHz/T respectively\cite{krzystek2015high}.

Applying a magnetic field perpendicular to the VO bond\cite{chicco2021controlled}, which corresponds to the $z$-direction for this structure, the coupling of each of the two spin levels of the electron to the eight spin levels of the nucleus through the hyperfine interaction results in the structure of the hybrid electron-nucleus spin energy levels as a function of the uniform magnetic field shown in Figure 2.

\begin{figure}[htp]
    \centering
    \includegraphics[width=8cm]{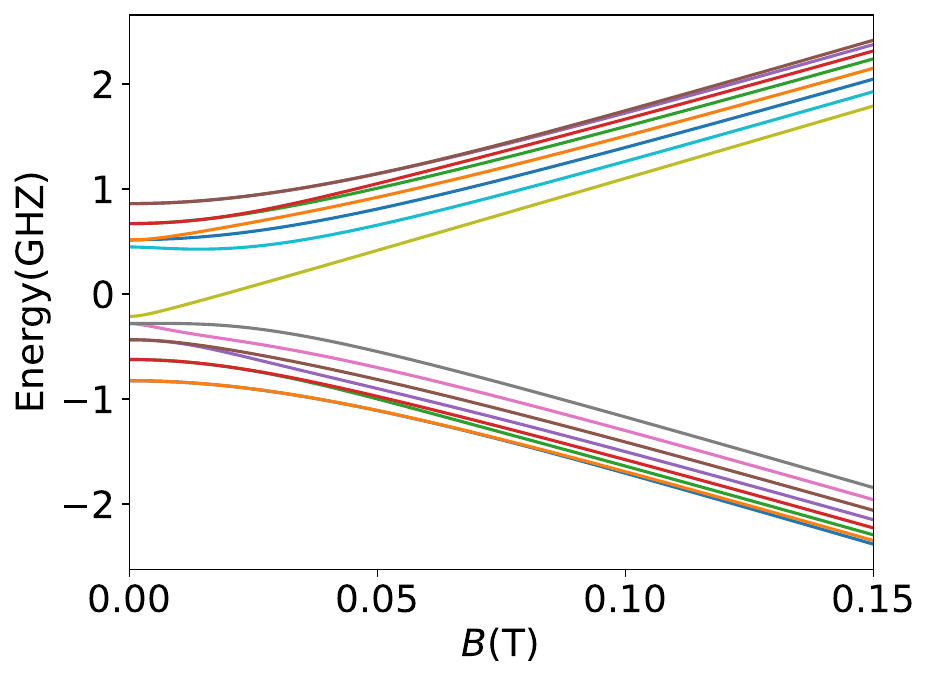}
        \label{fig:VOTPP_Energy}
    \caption{\textbf{Spin energy levels of VO(TPP).} The upper set of 8 energy levels corresponds to the 8 spin levels of the spin-7/2 nucleus coupled to the $m_s=1/2$ spin level of the electron. The lower set of 8 levels then corresponds to the nuclear levels coupled to the $m_s=-1/2$ spin level of the electron. Energy is shown as a function of the uniform magnetic field, $B$.}
\end{figure}

Experimentally, the measurement of $T_2$ is performed by detecting Hahn spin-echo signals and extracting $T_2$ from their decay. To detect these signals, the transition frequency of the qubit energy levels must allow only a change in $m_s$, not $m_I$\cite{yamabayashi2018scaling}. The qubit levels selected for the simulations in this work are therefore from the electron spin transition $|m_s = -1/2, m_I = -1/2\rangle\leftrightarrow|m_s = 1/2, m_I = -1/2\rangle$. The energy level splittings seen in Figure 2 show that all electron spin transitions with a fixed nuclear spin level have similar transition frequencies, making the choice of the qubit levels arbitrary, and the results independent of which specific electron spin transition is chosen for the qubit levels.

The surrounding spin bath of the qubit is split into contributions from the spins of other electrons, hydrogen nuclei, carbon nuclei, and nitrogen nuclei, with the carbon and nitrogen contributions considered to be negligible due to their lower populations and gyromagnetic ratios relative to the other contributing spins. In line with previous experimental studies of this system, the electron concentration of the bath is simulated at the diluted level of 2\% (44.65 mM) electron spin concentration\cite{yamabayashi2018scaling} to calculate a value of $T_2$ to compare with the nuclear contribution of the fully concentrated hydrogen nuclear spin bath. Simulations are conducted in the presence of a magnetic field of $B$ = 0.33 T, consistent with EPR measurements at the X-band frequency, and using convergent gCCE parameters. For the simulation of decoherence due to the electronic spin bath case, only clusters including up to 3 spins separated by no more than 40 \AA$ $ are considered, and only bath spins within a radius of 90 $\text{\AA}$ from the central one are included in the simulation. The profile for the electronic spin bath is averaged over 50 random realizations of the bath to account for the many possible distributions of the electron population inside the crystal that can occur at a concentration of 2\%. For the nuclear spin bath case, spin clusters of maximum size 2, separated by no more than 8 \AA$ $, and closer than 20 \AA$ $ from the central spin are instead considered. Supplementary Materials report a full discussion of how these parameters are obtained. 

The decoherence profiles of the qubit immersed in a bath of electrons and a bath of hydrogen nuclei are shown at the top and center of Figure 3. Fitting both decoherence profiles using a stretched exponential,
\begin{equation}
    \mathcal{L}(t) = e^{-\left(\frac{t}{T_2}\right)^\beta},
\end{equation}
shown in red on each plot, the electron bath case results in a coherence time of $T_2$ = 0.35 $\mu$s, whereas the coherence time in the hydrogen bath is $T_2$ = 10.88 $\mu$s, over an order of magnitude longer. Interestingly, if the electron is considered to be free instead of bound to the $^{51}$V nucleus, the predicted coherence time is $T_2$ = 0.17 $\mu$s for the electronic spin bath, a further underestimate, while a virtually identical value of $T_2$ = 11 $\mu$s is obtained for the hydrogen nuclear spin bath (see Supplementary Material for further discussion). The stretch factors calculated in each case are $\beta$ = 0.92 for the electronic spin bath, and $\beta$ = 2.2 for the hydrogen nuclear spin bath. The electronic spin bath stretch factor is close to 1, which can be understood as due to the rapid loss of coherence expected for Markovian dynamics or the averaging over multiple contributions with stretch factor larger than 1\cite{lunghi2019electronic}. The nuclear spin bath value for the stretch factor is in line with expectations as a value between 2 and 3 is expected when the decoherence is dominated by spin flip-flops between the nuclei in the bath\cite{lenz2017quantitative}. Finally, we also perform simulations on a deuterated crystal using the following converged simulation parameters: spin clusters of size 2 separated by a maximum distance of 6 $\text{\AA}$ and within a radius of 20 $\text{\AA}$ from the central spin. The decoherence profile at the bottom of Figure 3 produces a coherence time of 127 $\mu$s with a stretch factor of 1.69. 

The decoherence data for the VO(TPP) molecular qubit calculated theoretically through gCCE for each spin bath case and measured experimentally are summarized in Table II, with the coherence time of the spin bath that primarily contributes to the experimentally observed decoherence highlighted in bold.

\begin{table}[htp]
\begin{center}
    \label{table:VOTPP_results}
    \caption{Decoherence data for VO(TPP)}
    \begin{tabular}{ |p{2cm}|p{2cm}|p{2cm}|p{2cm}|  }
     \hline
     Sample & e bath & n bath & Exp\\
     \hline
     2\%-H & \textbf{0.35} $\mu$s & 10.88 $\mu$s & 1 $\mu$s\\
     2\%-D & -- &  127 $\mu$s & -- \\
     \hline
    \end{tabular}
\end{center}
\end{table}

For the experimentally tested sample, results show that the experimental value of 1 $\mu$s is closer to the $T_2$ of the electronic spin bath ($\approx$ 3$\times$ smaller) than the $T_2$ of the hydrogen spin bath ($\approx$ 10$\times$ larger), supporting the conclusion that the spin decoherence of the VO(TPP) molecular qubit at an electron spin concentration of 2\% is still dominated by spin-spin interactions among the electron bath spins rather than the spins of the hydrogen nuclei, which have been suspected to be the main contributors to decoherence in previous experiments carried out at these exact conditions.

\begin{figure}[htp]
    \centering
    \includegraphics[width=8cm]{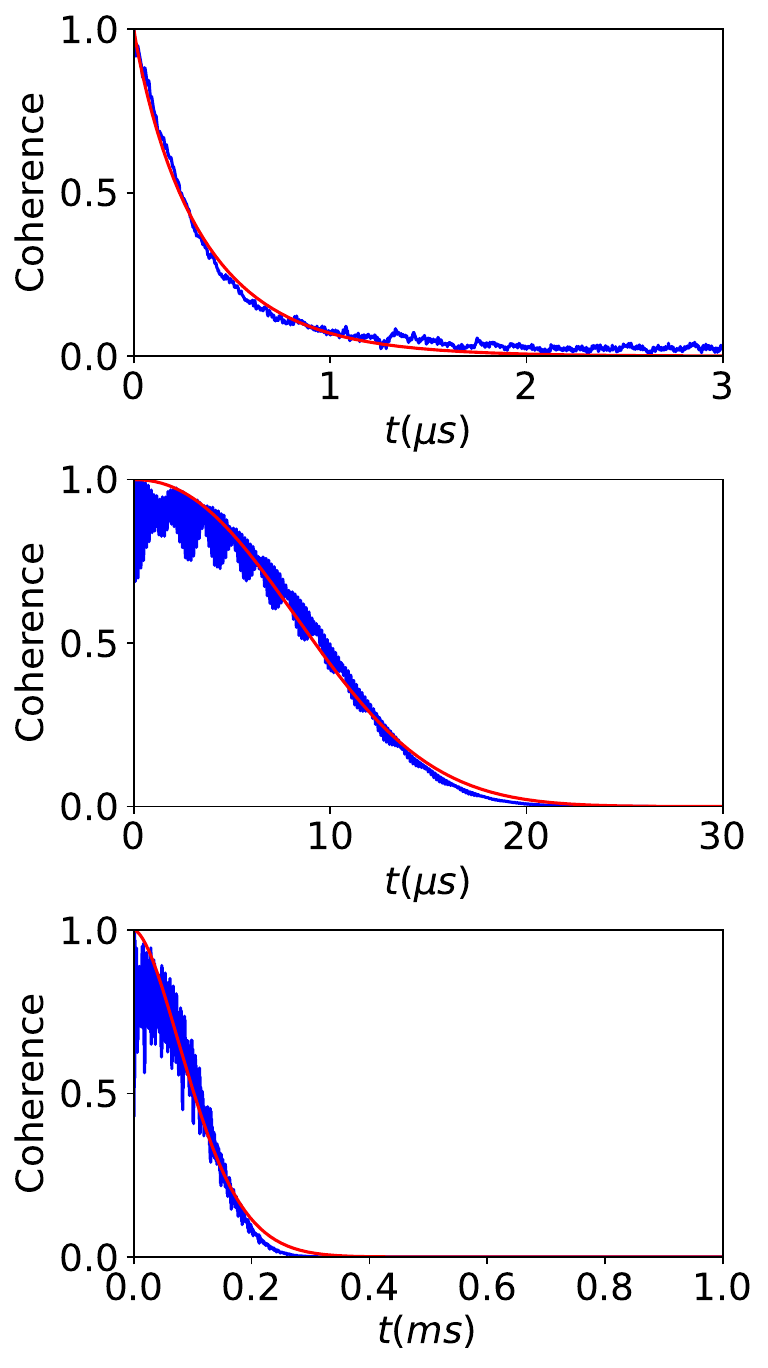}
    \label{fig:VOTPP_decoherence}
    \caption{\textbf{VO(TPP) decoherence profile.} Decoherence profiles of the $^{51}$V electron spin in a bath of electrons (top), a bath of hydrogen nuclear spins (center), and a bath of deuterium nuclear spins (bottom).}
\end{figure}

We then proceed to estimate the percentage of dilution at which a crossover occurs from the regime in which the electronic spin bath is the leading contribution to decoherence to the regime in which the hydrogen nuclear spin bath leads. To accomplish this, $T_2$ is calculated for electronic spin bath concentrations ranging from 2-50\%. The chosen range of concentrations has to be large enough to clearly capture the dependence of $T_2$ on 1/concentration\cite{zecevic1998dephasing,bar2012suppression, ye2019spin}. The specific range of 2-50\% used in this work is chosen to fully capture the relationship between $T_2$ and the concentration, with the upper limit of 50\% chosen because the decoherence dynamics become more susceptible to divergences at higher concentrations, making it difficult to obtain accurate coherence times. This relationship between $T_2$ and concentrations then enables a linear fit to be used in a log-log plot to extrapolate to lower concentrations. Using the fit shown in Figure 4, it is found that the crossover occurs at a concentration of 0.049\% (1.1 mM), in qualitative agreement with previously made predictions based on the simulation of decoherence in vanadyl-based molecular qubits\cite{lunghi2019electronic}. To check for consistency, the calculated coherence time at 2\% concentration using extrapolation is found to be 0.26 $\mu$s, which is in good agreement with the previously calculated value of 0.35 $\mu$s when considering the accuracy of the linear fit in Figure 4. 
\begin{figure}[htp]
    \centering
        \includegraphics[width=8cm]{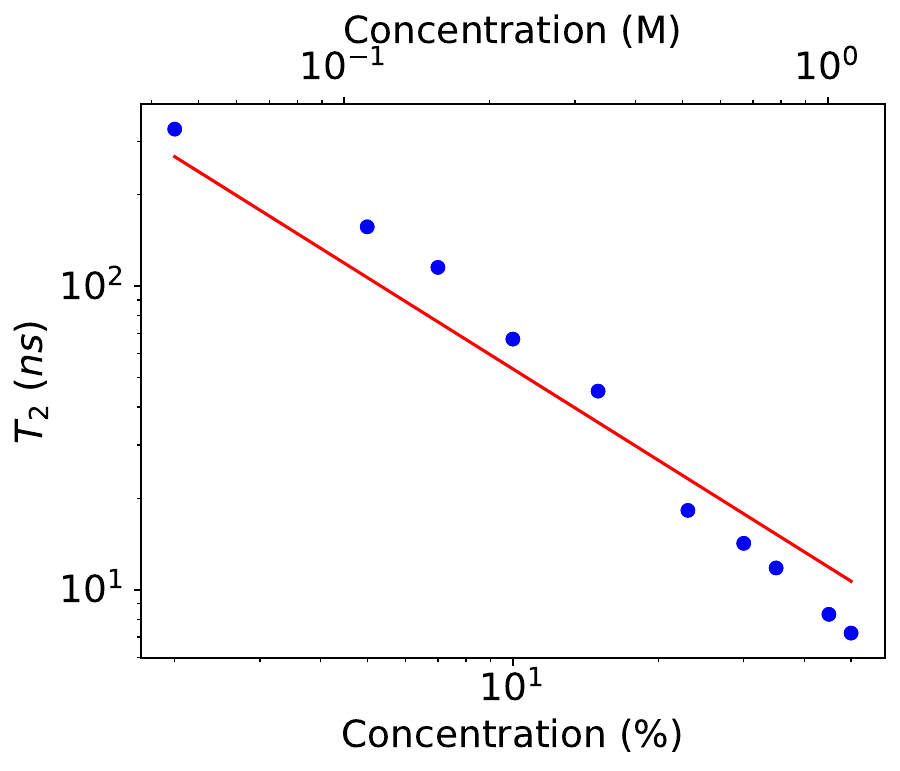}
        \label{fig:VOTPP_linear}
    \caption{\textbf{Decoherence vs magnetic dilution in VO(TPP).} Linear fit of $\log_{10}(T_2)$ against $\log_{10}$ of the concentration percentage of the electrons in the bath in order to extrapolate the value of $T_2$ at low concentrations.}
\end{figure}
\FloatBarrier
\noindent
With the aim of maximizing $T_2$ by not just diluting the electron spins to design a sample in the regime dominated by hydrogen spins, but instead in the regime dominated by deuterium spins, further dilution past 0.049\% (1.1 mM) is instead required. Extrapolating the fit in Figure 4 to $T_2$ = 127 $\mu$s, it is found to occur at electron bath spin concentrations in the range of 0.004\% (0.09 mM), a concentration achievable in experiments.

\subsection*{Decoherence in [Cu(mnt)$_2$]$^{2-}$}
\noindent
The [Cu(mnt)$_2$]$^{2-}$ (mnt$^{2-}$ = maleonitriledithiolate or 1,2-dicyanoethylene-1,2-dithiolate) molecular spin qubit contains a single unpaired spin-1/2 electron in the $d$-shell of the Cu(II) ion. The unit cell of the (PPh$_4$)$_2$[Cu(mnt)$_2$] crystal, obtained from its X-ray structure, is shown in Figure 5, and contains two copies of the [Cu(mnt)$_2$]$^{2-}$ molecular spin qubit and four copies of the [PPh$_4$]$^{2-}$ counter ions.
\begin{figure}[htp]
    \centering
    \includegraphics[width=8cm]{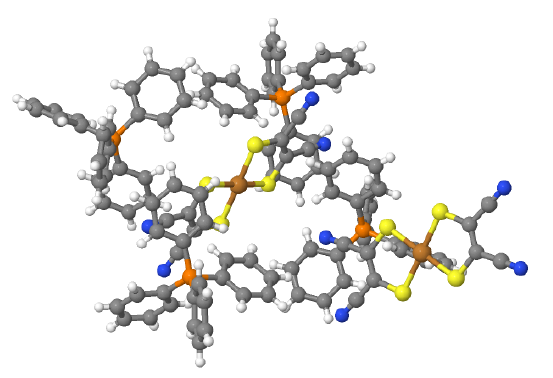}
        \label{fig:Cu}
\caption{\textbf{[Cu(mnt)$_2$]$^{2-}$ crystal structure.} The position of the atoms in the unit cell of (PPh$_4$)$_2$[Cu(mnt)$_2$] is reporteded with the color code: copper(brown), sulfur(yellow), nitrogen(blue), phosphorus(orange), carbon(dark grey), hydrogen(white).}
\end{figure}

The unpaired electron of [Cu(mnt)$_2$]$^{2-}$ interacts with the spin-3/2 $^{63}$Cu nucleus through a spin Hamiltonian of the same form as Eq (10). The spin Hamiltonian parameters in this case are the hyperfine tensor and electronic $g$-tensor with the components below,
\begin{table}[htp]
\begin{center}
    \label{table:Cu_H}
    \caption{Hamiltonian Parameters for [Cu(mnt)$_2$]$^{2-}$}
    \begin{tabular}{ |p{2cm}|p{2cm}|p{2cm}|p{2cm}|  }
     \hline
     $g_{\perp}$ & $g_{\parallel}$ & $A_{\perp}$ (MHz) & $A_{\parallel}$ (MHz)\\
     \hline
     2.0227 & 2.0925 & 118 & 500\\
     \hline
    \end{tabular}
\end{center}
\end{table}
measured from fits on ESE-detected Q-band EPR spectra in experiments\cite{bader2014room}. The quadrupolar coupling is then given by $p$ = 9.45 MHz, calculated from ORCA\cite{RN178}. The $g$-factor of the $^{63}$Cu nucleus is then 1.484 and its nuclear Bohr magneton is $\mu_N$ = 7.624 MHz/T.

With a magnetic field applied in the $z$-direction once again, the hyperfine coupling of each spin energy level of the unpaired electron to the 4 nuclear spin energy levels gives the change in energy of each spin level as a function of the uniform magnetic field across the crystal shown in Figure 6.
\begin{figure}[htp]
    \centering
    \includegraphics[width=8cm]{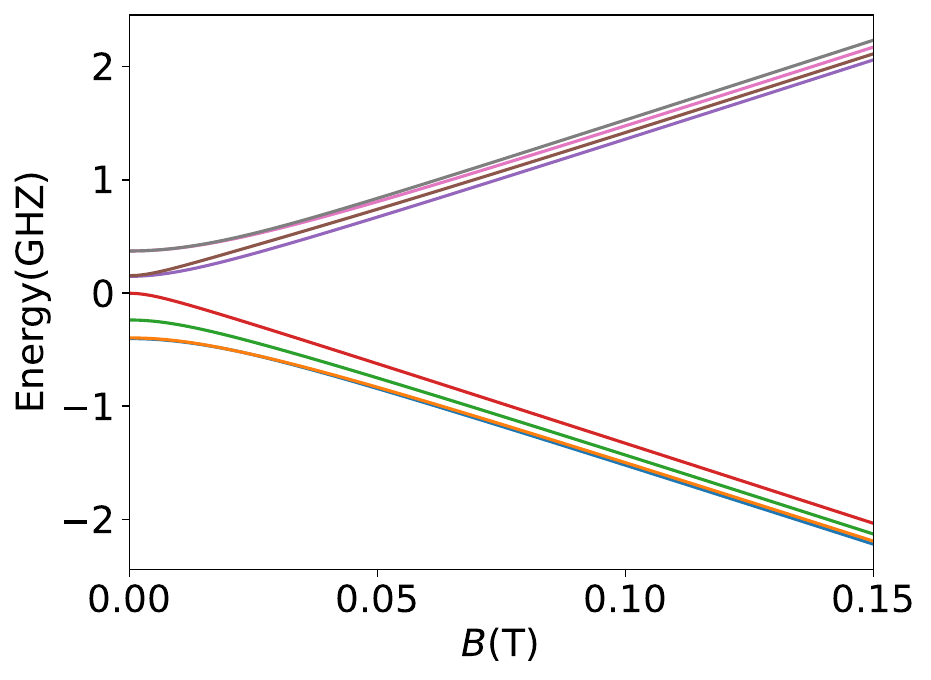}
        \label{fig:Cu_Energy}
    \caption{\textbf{Spin energy levels of [Cu(mnt)$_2$]$^{2-}$.} The upper set of 4 energy levels corresponds to the 4 spin levels of the spin-3/2 nucleus coupled to the $m_s=1/2$ spin level of the electron. The lower set of 4 levels then corresponds to the nuclear levels coupled to the $m_s=-1/2$ spin level of the electron. Energy is shown as a function of the uniform magnetic field, $B$. }
\end{figure}
As a result of the same reasoning explained in selecting the qubit levels for VO(TPP), the qubit levels for the [Cu(mnt)$_2$]$^{2-}$ molecular qubit are again chosen to be on the electron spin transition $|m_s=-1/2,m_I=-1/2\rangle\leftrightarrow|m_s=1/2,m_I=-1/2\rangle$.

The coherence time of [Cu(mnt)$_2$]$^{2-}$ has been experimentally studied both by diluting the electronic spin bath to a concentration of 0.001\% (0.015 mM), sample 0.001\%-H from now on, and by diluting it to a concentration of 0.01\% (0.15 mM) while simultaneously replacing hydrogen nuclear spins in the (PPh$_4$) counter-ions for deuterium nuclear spins\cite{bader2014room}, sample 0.01\%-D from now on, which carry a much weaker gyromagnetic ratio ($\gamma_{\text{D}}$ = 4.1065 rad$\cdot$ms$^{-1}\cdot$G$^{-1}$) and therefore are more weakly interacting than the hydrogen nuclear spins ($\gamma_{\text{H}}$ = 26.7522 rad$\cdot$ms$^{-1}\cdot$G$^{-1}$). Table IV contains the experimentally measured values of $T_2$ for both samples, 0.001\%-H and 0.01\%-D. For the 0.001\%-H sample, a coherence time of 9.23 $\mu$s is measured, while for the 0.01\%-D sample, the coherence time increases to 68 $\mu$s.
\begin{table}[htp]
    \begin{center}
    \label{table:Cu}
    \caption{Decoherence data for [Cu(mnt)$_2$]$^{2-}$}
    \begin{tabular}{ |p{2cm}|p{2cm}|p{2cm}|p{2cm}| }
     \hline
     Sample & e bath & n bath & Exp \\
     \hline
     0.001\%-H & 797 $\mu$s & \textbf{8.6} $\mu$s & 9.23 $\mu$s \\
     0.01\%-D & \textbf{80} $\mu$s & 100 $\mu$s & 68 $\mu$s \\     
     \hline
    \end{tabular}
    \end{center}
\end{table}
With experiments having been carried out at the more diluted concentrations of 0.01\% and 0.001\%, a direct simulation of the decoherence as in the case of the VO(TPP) molecular qubit is made more difficult due to the large increase in the number of bath realizations required to average over to account for the far more sparsely populated electronic spin bath. The method of calculating $T_2$ for a range of concentrations and extrapolating to inaccessible concentration percentages is therefore used to analyze all behavior in relation to the electronic spin bath. Using convergent parameters calculated for gCCE, that is, gCCE order 2, bath radius 25 $\text{\AA}$, and maximum bath spin dipole-dipole interaction distance 8 $\text{\AA}$ (more details in Supplementary Material), the decoherence profile shown at the top of Figure 7 produces a coherence time of $T_2$ = 8.6 $\mu$s and a stretch factor of $\beta$ = 2.2 for the qubit immersed in a hydrogen spin bath. The electronic spin bath extrapolation shown in Figure 8 then predicts a coherence time of 797 $\mu$s at the experimentally used concentration. The value of $T_2$ for the hydrogen spin bath is very clearly much closer to the value of 9.23 $\mu$s measured in experiments\cite{bader2014room}, proving that the decoherence of this sample is dominated by the hydrogen nuclear spin bath, rather than the 0.001\% electronic spin bath. The coherence time of 8.6 $\mu$s for this sample is therefore highlighted in bold in Table IV to mark the hydrogen nuclear spin bath as the main source of decoherence. 

\begin{figure}[htp]
    \centering
    \includegraphics[width=8cm]{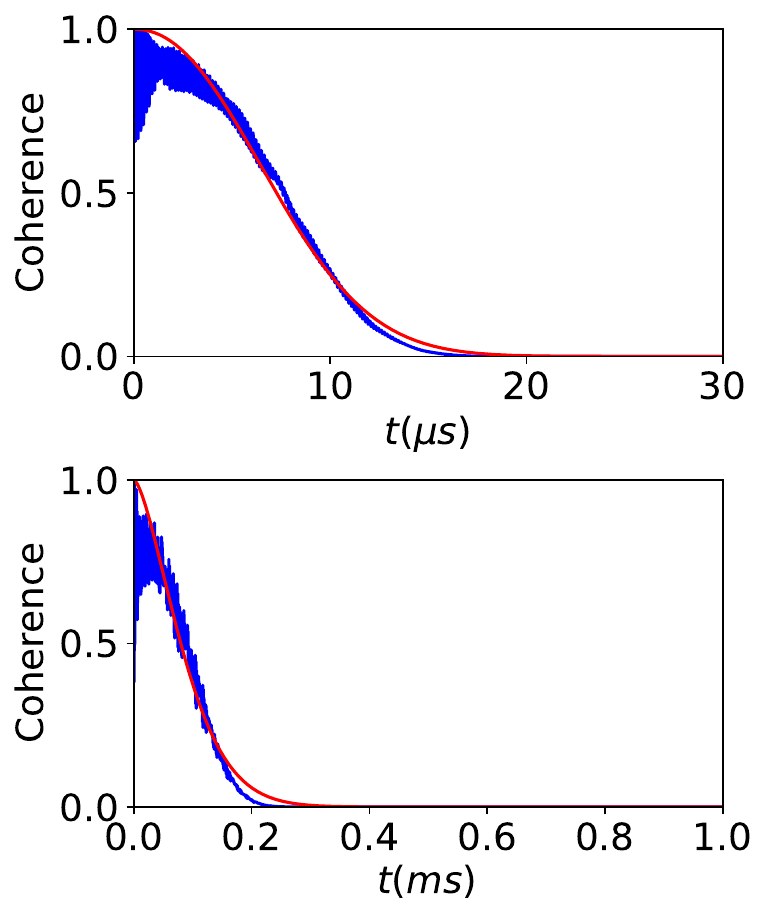}
    \label{fig:Cu_decoherence}
    \caption{\textbf{[Cu(mnt)$_2$]$^{2-}$ decoherence profile.} Decoherence profiles of the $^{63}$Cu electron spin in a bath of hydrogen nuclear spins (top) and a bath of deuterium nuclear spins (bottom).}
\end{figure}

Performing now gCCE simulations for a fully deuterated sample with convergent simulation parameters of gCCE order 2, bath radius 20 $\text{\AA}$, and maximum bath spin dipole-dipole interaction distance 8 $\text{\AA}$, the decoherence dynamics shown at the bottom of Figure 7 produces a coherence time of $T_2$ = 0.1 ms and a stretch factor of $\beta$ = 1.5 for the qubit surrounded by deuterium nuclei, with identical values for both of these parameters also calculated for the free electron (further detail in the Supplementary Material). Extrapolating from the linear fit in Figure 8 for the electronic spin bath then provides a value of $T_2$ = 80 $\mu$s. Although both baths result in similar values, the one obtained for the electron bath is in very close agreement with the experimentally observed $T_2$, pointing to the latter as the main limit to coherence for the 0.01\%-D sample.

Extending the theoretical predictions from gCCE, the extrapolation of the linear fit predicts an electron spin concentration that gives a coherence time comparable to that of the hydrogen spin bath, namely 0.093\% (1.4 mM).
\begin{figure}[htp]
    \centering
    \includegraphics[width=8cm]{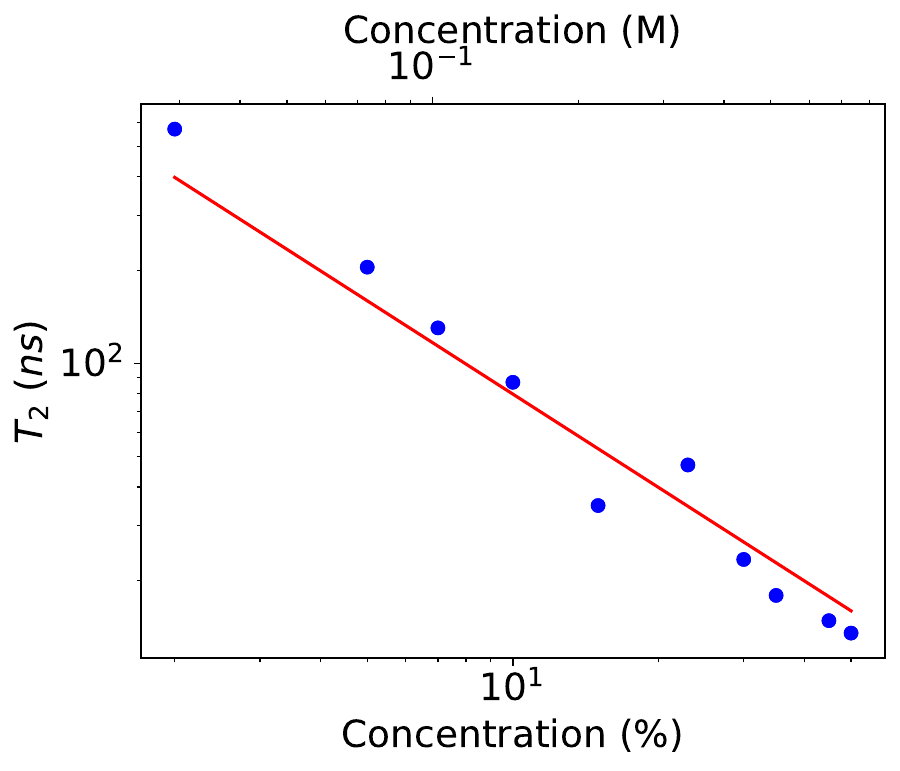}
    \label{fig:Cu_linear}
    \caption{\textbf{Decoherence vs magnetic dilution in [Cu(mnt)$_2$]$^{2-}$.} Linear fit of $\log_{10}(T_2)$ against $\log_{10}$ of the concentration percentage of the electrons in the bath in order to extrapolate the value of $T_2$ at low concentrations.}
\end{figure}
To extend $T_2$ to the larger upper limit dictated by the deuterium nuclear spin bath, Figure 8 predicts a crossover point at an electronic spin bath concentration of 0.008\% (0.1 mM), again very similar to the crossover point predicted for VO(TPP).

Up to this point, a quantitative breakdown of the contributions from each individual spin bath has been performed, with there now being a clear outline for how to extend $T_2$ towards millisecond timescales through electron spin dilution and deuteration of the hydrogen spins. One final aspect of these experiments that has yet to be explicitly considered is the role of a dynamical decoupling (DD) pulse sequence. All results produced so far use the standard Hahn echo pulse sequence, ($\pi$/2)$_x-t/2-(\pi)_y-t/2$. This pulse sequence has the effect of first initializing the qubit spin in an equal superposition state, then halfway through the time evolution of the qubit a refocusing $\pi$-pulse inverts the qubit spin in order to eliminate static inhomogeneous contributions to decoherence, such as the effect of a non-uniform magnetic field across the sample. While the Hahn echo pulse sequence is successful in extending the coherence time of a spin from the value measured in Free Induction Decay (FID), the robustness of these pulses against errors\cite{shim2012robust}, such as fluctuations of the microwave field, and advancements in experimental equipment allow for the application of thousands of refocusing $(\pi)_y$ pulses in a Carr-Purcell-Meiboom-Gill (CPMG) pulse sequence\cite{ronczka2012optimization} of the form $(\pi/2)_x-\{t/2-(\pi)_y-t/2\}_n$, where $n$ is the number of refocusing pulses.

To theoretically study the effect of extended pulse sequence prolonging coherence times in molecular qubits, we now model the spin of the [Cu(mnt)$_2$]$^{2-}$ molecular qubit as that of a free electron for the purpose of a smaller computational workload. Based on previously made comments on the results in this work, this further approximation should still produce highly accurate results for a bath of nuclear spins, although it may lead to slightly less accurate results for the electronic spin bath due to factors such as enhanced relaxation effects, the statistical distribution of the bath spins, and the perturbative arguments on which CCE is built being pushed to their limit.

\begin{figure}[h!]
    \centering
    \includegraphics[width=8cm]{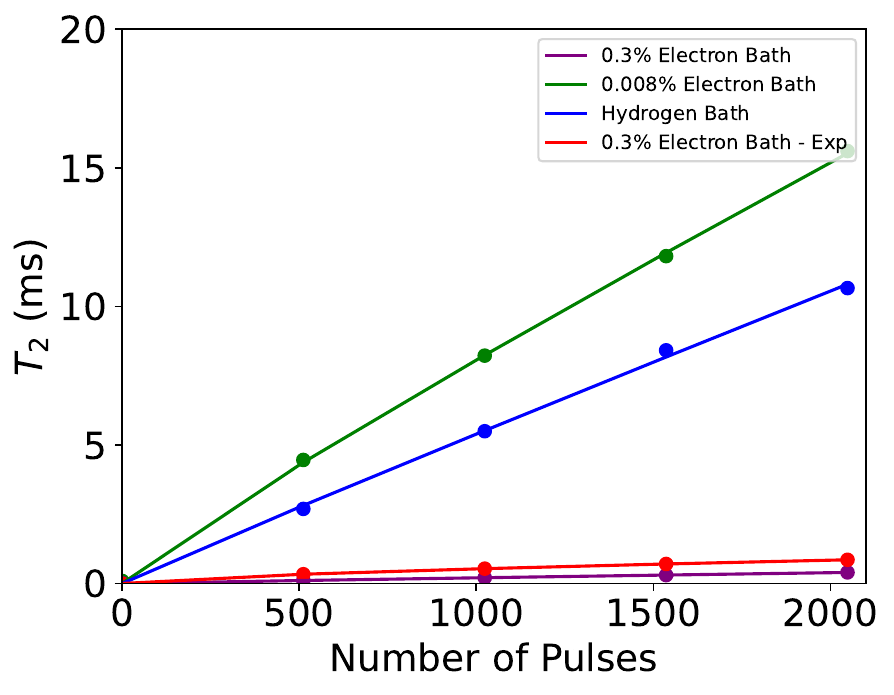}
    \label{fig:Cu_pulses}
    \caption{\textbf{Coherence under dynamical decoupling.} Scaling of the coherence time of the [Cu(mnt)$_2$]$^{2-}$ molecular qubit with the number of refocusing $\pi$ pulses in an electronic spin bath at 0.3\% concentration (theory - purple line, experiment - red line), an electron bath at 0.008\% concentration (green line), and a hydrogen spin bath (blue line).}
\end{figure}

Extended pulse sequences have previously been applied to the [Cu(mnt)$_2$]$^{2-}$ molecular qubit in experiments by Dai et al.\cite{dai2021experimental}, resulting in a prolonged coherence time of 1.4 ms. While this extension of the coherence time to millisecond timescales is an impressive result, this result was measured at an electronic spin bath concentration of 0.3\%, still in the regime we predict to be dominated by electron spin decoherence and therefore possible to improve upon through further dilution of the sample. To study the level of improvement that can be expected of $T_2$ when combining the effectiveness of DD pulse sequences with the effectiveness of diluting the electron spin bath to the levels suggested in this work, CPMG pulse sequences up to 2048 pulses are applied in the simulation of decoherence in the hydrogen spin bath, an electronic spin bath diluted to 0.008\% (the expected crossover point with the deuterium spin bath), and an electronic spin bath diluted to 0.3\% to compare against the experimental data. Informed by previous studies on how $T_2$ scales with the number of CPMG pulses\cite{bar2013solid,dai2021experimental}, the data in all cases is fitted using a power law, $T_2(n)$ = $T_2^0n^p$, with $T_2^0$ corresponding to the coherence time measured for a single refocusing pulse. All fitting parameters for each case are gathered in Table V.

\begin{table}[htp]
    \begin{center}
    \label{table:Power_law}
    \caption{Power Law Parameters for [Cu(mnt)$_2$]$^{2-}$}
    \begin{tabular}{ |p{1.5cm}|p{1.25cm}|p{0.75cm}|p{1.25cm}|p{0.75cm}|p{1.25cm}|p{0.75cm}| }
     \hline
     Sample & \multicolumn{2}{p{1cm}|}{e bath} & \multicolumn{2}{p{1cm}|}{n bath} & \multicolumn{2}{p{1cm}|}{Exp} \\
     \hline
      & $T_2^0$ ($\mu$s) & $p$ & $T_2^0$ ($\mu$s) & $p$ & $T_2^0$ ($\mu$s) & $p$ \\
     \hline
     0.3\%-H & 0.38 & 0.91 & 6.7 & 0.97 & 5.2 & 0.67 \\
     0.008\%-H & 14.69 & 0.91 & 6.7 & 0.97 & -- & -- \\
     \hline
    \end{tabular}
    \end{center}
\end{table}

\noindent
Fitted parameters for the experimental data are $T_2^0$ = 5.2 $\mu$s for the single pulse coherence time, and a power of 0.67 for setting the rate at which $T_2$ increases with respect to the number of pulses. This rate of increase is in good agreement with other studies on the change in $T_2$ when using large DD pulse sequences\cite{bar2013solid}. For the theoretical predictions made using gCCE, this rate of increase is described by the large power of 0.91, in disagreement with what appears to be the general consensus, and therefore warranting further investigation for the cause of computational methods predicting a greater rate of increase in the coherence time as more DD pulses are applied. There is also a noticeable disagreement between the experimental single pulse coherence time of 5.2 $\mu$s and the theoretical calculation of 0.38 $\mu$s, with the experimental measurement in fact being closer to the 6.7 $\mu$s single pulse coherence time of the hydrogen spin bath. However, we believe that this is not a result which contradicts our previous conclusions on which spin bath dominates decoherence in this sample, but rather is a consequence of the difficulties of approximating the electronic spin bath mentioned previously and are further elaborated on in the Discussion and Conclusions. This belief is further justified by the close agreement of the experimental and theoretical coherence times for the 0.03\% concentrated electronic spin bath shown in Figure 9. If the decoherence were instead dominated by nuclear spins, which would have to be hydrogen spins for this sample, $T_2$ should scale resembling the blue curve in Figure 9 and behave very differently to the experimentally measured data. It is also worth emphasizing that the single pulse coherence time for the hydrogen spin bath in these results is calculated through fitting the change in $T_2$ with a power law, whereas previously discussed results directly calculate a coherence time of 8.6 $\mu$s using the decoherence dynamics for single refocusing pulse in the hydrogen spin bath, slightly below the experimentally measured 9.23 $\mu$s\cite{bader2014room}, which should both be considered as more accurate and true to reality than a fitted parameter. The clearest evidence that the hydrogen spin bath is playing no significant role in the experimental change in $T_2$ as a function of the number of pulses in the CPMG sequence is that the rate of increase in $T_2$ is far greater than is seen experimentally, with the theoretical data calculating a power of 0.97, making the increase almost linear. At the largest number of pulses considered, 2048, the coherence time as a result of spin flip-flops of hydrogen nuclear spins is extended to 10.66 ms, a significant increase on what was measured experimentally. Pushing the dilution even further, to 0.008\%, where the electronic spin bath contribution to decoherence is expected to cross over with the deuterium spin contribution if the sample is deuterated, a further increase in $T_2$ under the action of DD pulses is predicted. The single pulse coherence time is fitted as 14.69 $\mu$s, the largest value so far, which is expected as we have now diluted beyond the concentration where the electronic spin bath and hydrogen spin bath will contribute equally. The rate at which $T_2$ increases is then identical to that of the theoretically calculated 0.3\% electronic spin bath results, with a power of 0.91. These computational results are therefore saying that the rate of increase of $T_2$ is dependent only the spin type of the dominant spin bath, and not its concentration. After applying a 2048 pulse CPMG pulse sequence $T_2$ is extended to 15.61 ms, over a factor of 11 greater than the experimentally measured result, emphasizing the effectiveness of combining both dilution with the application of large DD pulse sequences to dramatically increase $T_2$.

The upper limit of 2048 pulses is chosen in this work for a direct comparison with experiments\cite{dai2021experimental} and already pushes $T_2$ not only towards millisecond timescales, but towards achieving values on the order of tens of milliseconds. Assuming minimal error in the pulses, the exact upper limit on the number of pulses that can be applied as part of the CPMG sequence is only limited by being able to apply the pulses faster than the bath auto-correlation time, which characterizes how long it is possible to retrieve the coherence from the spin bath before it is irreversibly dissipated away. The number of pulses that can be applied in DD pulse sequences is certainly beyond the number considered in this work, with previous experiments studying decoherence in NV centers having applied up to 8192 pulses\cite{bar2013solid}. Once a sufficiently high number of pulses is applied, it is expected that the contribution to decoherence due to pure dephasing will essentially vanish, i.e. $T_2^*\rightarrow\infty$, and $T_2$ will then become limited by spin relaxation, $T_2$ = $2T_1$, with the experiments of Dai et al.\cite{dai2021experimental} finding that $T_1$ reaches 25 ms for this molecular qubit at 8 K, indicating that this upper limit on $T_2$ is still far from being reached.

\section*{Discussion and Conclusions}
\noindent
Crystals of magnetic molecules have been used as benchmark systems for the development of magnetic resonance models for decades and are recently receiving a resurgence of interest in the context of quantum information science. In both these contexts, developing quantitative, predictive, computational models is key. On the one hand, it provides an ultimate testbed for the theoretical models used to interpret experiments, but also provides microscopic insights on spin dynamics hardly achievable through measurements alone.

Here we have used CCE to explore the limits of spin-spin dynamics on electron molecular spin coherence. Importantly, differently from most previous attempts, we have explored the effect of both hydrogen atoms' nuclear spins and the residual electron spins invariably present for finite values of magnetic dilution. 

Our results clearly show that accounting for the bath of electron spins is key to correctly reinterpret literature results that would otherwise present inconsistencies. In the case of VO(TPP) we were able to show that the level of dilution included in the study of Yamabayashi et al.\cite{yamabayashi2018scaling} and subsequent ones, does not make it possible to observe the limit of coherence imposed by hydrogen atoms' nuclear spins, as generally claimed. On the other hand, we confirm the conclusion of Bader et al.\cite{bader2014room} on hydrogen nuclear spins limiting the coherence of [Cu(mnt)$_2$]$^{2-}$, but only for non-deuterated samples at 0.001\% concentration. In the case of deuterated samples, a too high electron spin concentration was used (0.01\%), preventing them from measuring the limit on $T_2$ imposed by deuterium's nuclear spins.

Our numerical simulations also provide clear indications on the level of dilution necessary to observe the limiting contribution of hydrogen and deuterium, pointing to levels of around 0.01\% (1 mM), and 0.001\% (0.1 mM) of electron spin concentration, respectively. It must be stressed that these target concentrations at which the electronic spin bath contribution to decoherence will be eliminated relative to the effect of nuclear spins in the bath are dependent on the nuclear spin concentrations, although the latter is likely consistent across the majority of molecular crystals and in agreement with findings for organic radicals\cite{zecevic1998dephasing,soetbeer2018dynamical}.

The very high level of dilution required to observe the limiting contribution of deuterium is rarely used in experiments, with two notable exceptions being the works of Bader et al.\cite{bader2014room}, already discussed, and Zadrozny et al.\cite{zadrozny2015millisecond}, both employing molecular qubits with proton-free organic ligands. Interestingly, in the latter work of Zadrozny et al.\cite{zadrozny2014multiple} a high dilution of 0.01 mM was only used for measurements in proton-free frozen solvents, while deuterated environments were measured at much higher electron spin concentrations, masking the effect of deuteration and making this limit yet to be observed experimentally in coordination compounds, to the best of our knowledge.

Our simulations indeed suggest that even in deuterated crystals $T_2$ can achieve very long values of 0.1 ms, well comparable to the values of 0.7 ms reported by Zadrozny for a proton-free molecular qubits in proton-free frozen solvent\cite{zadrozny2015millisecond}. This finding has important consequences on the design strategies for novel molecular qubits, shifting away attention to coming up with new, stable, proton-free organic ligands. Indeed, once a deuteration has been achieved or dynamical decoupling is employed, and the electronic spin is diluted to $\sim$ 0.01 mM concentrations, $T_2$ would once again become dominated by spin-phonon interactions already at a few K temperature, shifting the attention on extending the $T_1$ and $T_2$ due to the latter contributions\cite{mariano2025role}.

Now that we have established the limits of molecular spin coherence due to both electronic and nuclear spins, we are in a perfect position to compare molecular systems with other spin-based quantum platforms, most notably solid-state color defects such as NV centers and other vacancies in 3D and 2D materials. Interestingly, the limits of $T_2$ for NV centers at low temperature are in the range of 1 ms, as measured by a Hanh spin-echo pulse sequence\cite{bar2013solid}, less than one order of magnitude higher than what is estimated here for a molecular qubit like VO(TPP) in its own deuterated molecular crystal. The comparison becomes even more favorable for molecular spins when nanodiamonds or shallow defects are considered. The latter, necessary for any practical application of NV centers as qubits, exhibit values of $T_2$ on the order of 50 $\mu$s and only through complex material processing\cite{de2020temperature,wood2022long}, making molecular qubits a very competitive platform, where the same level of coherence time as solid-state qubits is achieved, but with the advantage of having a flexible, chemically tunable, easy to manipulate and fabricate physical platform.

Interestingly, the long values of $T_2$ in deeply buried color centers are consistently achieved not only thanks to the low concentration of nuclear spins present in bulk diamond, but also thanks to the very high levels of dilution of electronic spin centers employed. Thanks to the much lower detection limits of optically addressable magnetic resonance, compared to ordinary EPR, these systems are often measured at nM concentrations\cite{bar2013solid}, orders of magnitude lower than what is commonly employed for pulsed-EPR and well below the threshold we individuate to remove the contribution of electron-electron interactions. Efforts to produce optically addressable molecular qubits are well on their way\cite{kuppusamy2024spin}, suggesting that achieving unprecedented levels of coherence in molecular qubits is a possible reality.

The level of accuracy achieved by numerical simulations requires some commentary. In particular, whilst the calculation of $T_2$ due to the nuclear spin baths seems to always be very close to the experimental values, a slightly larger span of errors is observed when $T_2$ is attributed to electronic spin baths. We interpret these findings by observing that our simulations isolate a single central electronic spin, leaving the bath completely unaffected and in its thermal equilibrium state. This situation is not necessarily met in practice during experiments, as the microwave pulses will excite multiple spins at the same time. This is particularly true for electronic spin baths, and more so at high concentrations, where multiple spins would simultaneously be flipped during an excitation pulse. Interestingly, the deviations observed for electronic spin baths seem to be consistently reduced by considering the central electron spin bound to its own nuclear spin, which arguably provides a more realistic description of the system and more accurately accounts for spin-bath detuning effects.


In conclusion, we have here applied numerical simulations to the study of spin decoherence in molecular crystals taking into account both the effect of nuclear and electronic spin baths. Our results show the key role of electron spins in interpreting the values of molecular spin coherence reported in literature and point to values of $T_2$ approaching ms for highly-diluted deuterated samples or even exceeding 10 ms if dynamical decoupling is employed. These values are competitive with any other spin-based quantum physical platform and point to a potential prominent role of magnetic molecules for the field of quantum information science in the years to come.


\section*{Author Contributions}\noindent
A.L. and V.B. acknowledge funding from the European Research Council (ERC) under the European Union’s Horizon 2020 research and innovation programme (grant agreement No. [948493]) and C.R. from Taighde Éireann - Research Ireland through the Government of Ireland Postgraduate Scholarship. Computational resources were provided by Trinity College Research IT and the Irish Centre for High-End Computing (ICHEC).

\section*{Author Contributions}
\noindent
A.L. conceived the project and supervised its execution. All authors have contributed to the development and testing of the code used for running simulations. C.R. performed all simulations presented in the manuscript. C.R. and A.L. wrote the manuscript with input from all other authors.

\section*{Conflict of Interest}
\noindent
The authors declare that they have no competing interests.


\bibliographystyle{naturemag}
\bibliography{biblio}

\begin{thebibliography}{10}
\expandafter\ifx\csname url\endcsname\relax
  \def\url#1{\texttt{#1}}\fi
\expandafter\ifx\csname urlprefix\endcsname\relax\def\urlprefix{URL }\fi
\providecommand{\bibinfo}[2]{#2}
\providecommand{\eprint}[2][]{\url{#2}}

\bibitem{moreno2018molecular}
\bibinfo{author}{Moreno-Pineda, E.}, \bibinfo{author}{Godfrin, C.}, \bibinfo{author}{Balestro, F.}, \bibinfo{author}{Wernsdorfer, W.} \& \bibinfo{author}{Ruben, M.}
\newblock \bibinfo{title}{Molecular spin qudits for quantum algorithms}.
\newblock \emph{\bibinfo{journal}{Chemical Society Reviews}} \textbf{\bibinfo{volume}{47}}, \bibinfo{pages}{501--513} (\bibinfo{year}{2018}).

\bibitem{coronado2020molecular}
\bibinfo{author}{Coronado, E.}
\newblock \bibinfo{title}{Molecular magnetism: from chemical design to spin control in molecules, materials and devices}.
\newblock \emph{\bibinfo{journal}{Nature Reviews Materials}} \textbf{\bibinfo{volume}{5}}, \bibinfo{pages}{87--104} (\bibinfo{year}{2020}).

\bibitem{wasielewski2020exploiting}
\bibinfo{author}{Wasielewski, M.~R.} \emph{et~al.}
\newblock \bibinfo{title}{Exploiting chemistry and molecular systems for quantum information science}.
\newblock \emph{\bibinfo{journal}{Nature Reviews Chemistry}} \textbf{\bibinfo{volume}{4}}, \bibinfo{pages}{490--504} (\bibinfo{year}{2020}).

\bibitem{lavroff2021recent}
\bibinfo{author}{Lavroff, R.~H.} \emph{et~al.}
\newblock \bibinfo{title}{Recent innovations in solid-state and molecular qubits for quantum information applications} (\bibinfo{year}{2021}).

\bibitem{gaita2019molecular}
\bibinfo{author}{Gaita-Ari{\~n}o, A.}, \bibinfo{author}{Luis, F.}, \bibinfo{author}{Hill, S.} \& \bibinfo{author}{Coronado, E.}
\newblock \bibinfo{title}{Molecular spins for quantum computation}.
\newblock \emph{\bibinfo{journal}{Nature chemistry}} \textbf{\bibinfo{volume}{11}}, \bibinfo{pages}{301--309} (\bibinfo{year}{2019}).

\bibitem{fursina2023toward}
\bibinfo{author}{Fursina, A.~A.} \& \bibinfo{author}{Sinitskii, A.}
\newblock \bibinfo{title}{Toward molecular spin qubit devices: Integration of magnetic molecules into solid-state devices}.
\newblock \emph{\bibinfo{journal}{ACS Applied Electronic Materials}} \textbf{\bibinfo{volume}{5}}, \bibinfo{pages}{3531--3545} (\bibinfo{year}{2023}).

\bibitem{yang2016quantum}
\bibinfo{author}{Yang, W.}, \bibinfo{author}{Ma, W.-L.} \& \bibinfo{author}{Liu, R.-B.}
\newblock \bibinfo{title}{Quantum many-body theory for electron spin decoherence in nanoscale nuclear spin baths}.
\newblock \emph{\bibinfo{journal}{Reports on Progress in Physics}} \textbf{\bibinfo{volume}{80}}, \bibinfo{pages}{016001} (\bibinfo{year}{2016}).

\bibitem{mondal2023spin}
\bibinfo{author}{Mondal, S.} \& \bibinfo{author}{Lunghi, A.}
\newblock \bibinfo{title}{Spin-phonon decoherence in solid-state paramagnetic defects from first principles}.
\newblock \emph{\bibinfo{journal}{npj Computational Materials}} \textbf{\bibinfo{volume}{9}}, \bibinfo{pages}{120} (\bibinfo{year}{2023}).

\bibitem{hahn1950spin}
\bibinfo{author}{Hahn, E.~L.}
\newblock \bibinfo{title}{Spin echoes}.
\newblock \emph{\bibinfo{journal}{Physical review}} \textbf{\bibinfo{volume}{80}}, \bibinfo{pages}{580} (\bibinfo{year}{1950}).

\bibitem{bar2013solid}
\bibinfo{author}{Bar-Gill, N.}, \bibinfo{author}{Pham, L.~M.}, \bibinfo{author}{Jarmola, A.}, \bibinfo{author}{Budker, D.} \& \bibinfo{author}{Walsworth, R.~L.}
\newblock \bibinfo{title}{Solid-state electronic spin coherence time approaching one second}.
\newblock \emph{\bibinfo{journal}{Nature communications}} \textbf{\bibinfo{volume}{4}}, \bibinfo{pages}{1743} (\bibinfo{year}{2013}).

\bibitem{gottscholl2021spin}
\bibinfo{author}{Gottscholl, A.} \emph{et~al.}
\newblock \bibinfo{title}{Spin defects in hbn as promising temperature, pressure and magnetic field quantum sensors}.
\newblock \emph{\bibinfo{journal}{Nature communications}} \textbf{\bibinfo{volume}{12}}, \bibinfo{pages}{4480} (\bibinfo{year}{2021}).

\bibitem{liu2022spin}
\bibinfo{author}{Liu, W.} \emph{et~al.}
\newblock \bibinfo{title}{Spin-active defects in hexagonal boron nitride}.
\newblock \emph{\bibinfo{journal}{Materials for Quantum Technology}} \textbf{\bibinfo{volume}{2}}, \bibinfo{pages}{032002} (\bibinfo{year}{2022}).

\bibitem{melnikov2004quantum}
\bibinfo{author}{Melnikov, D.~V.} \& \bibinfo{author}{Chelikowsky, J.~R.}
\newblock \bibinfo{title}{Quantum confinement in phosphorus-doped silicon nanocrystals}.
\newblock \emph{\bibinfo{journal}{Physical review letters}} \textbf{\bibinfo{volume}{92}}, \bibinfo{pages}{046802} (\bibinfo{year}{2004}).

\bibitem{jeong2010spin}
\bibinfo{author}{Jeong, M.} \emph{et~al.}
\newblock \bibinfo{title}{Spin coherence time t 2 in metallic p-doped si at very low temperature}.
\newblock \emph{\bibinfo{journal}{Journal of Low Temperature Physics}} \textbf{\bibinfo{volume}{158}}, \bibinfo{pages}{659--665} (\bibinfo{year}{2010}).

\bibitem{bader2014room}
\bibinfo{author}{Bader, K.} \emph{et~al.}
\newblock \bibinfo{title}{Room temperature quantum coherence in a potential molecular qubit}.
\newblock \emph{\bibinfo{journal}{Nature communications}} \textbf{\bibinfo{volume}{5}}, \bibinfo{pages}{5304} (\bibinfo{year}{2014}).

\bibitem{zadrozny2015millisecond}
\bibinfo{author}{Zadrozny, J.~M.}, \bibinfo{author}{Niklas, J.}, \bibinfo{author}{Poluektov, O.~G.} \& \bibinfo{author}{Freedman, D.~E.}
\newblock \bibinfo{title}{Millisecond coherence time in a tunable molecular electronic spin qubit}.
\newblock \emph{\bibinfo{journal}{ACS central science}} \textbf{\bibinfo{volume}{1}}, \bibinfo{pages}{488--492} (\bibinfo{year}{2015}).

\bibitem{atzori2016room}
\bibinfo{author}{Atzori, M.} \emph{et~al.}
\newblock \bibinfo{title}{Room-temperature quantum coherence and rabi oscillations in vanadyl phthalocyanine: toward multifunctional molecular spin qubits}.
\newblock \emph{\bibinfo{journal}{Journal of the American Chemical Society}} \textbf{\bibinfo{volume}{138}}, \bibinfo{pages}{2154--2157} (\bibinfo{year}{2016}).

\bibitem{lunghi2019phonons}
\bibinfo{author}{Lunghi, A.} \& \bibinfo{author}{Sanvito, S.}
\newblock \bibinfo{title}{How do phonons relax molecular spins?}
\newblock \emph{\bibinfo{journal}{Science advances}} \textbf{\bibinfo{volume}{5}}, \bibinfo{pages}{eaax7163} (\bibinfo{year}{2019}).

\bibitem{garlatti2023critical}
\bibinfo{author}{Garlatti, E.} \emph{et~al.}
\newblock \bibinfo{title}{The critical role of ultra-low-energy vibrations in the relaxation dynamics of molecular qubits}.
\newblock \emph{\bibinfo{journal}{Nature Communications}} \textbf{\bibinfo{volume}{14}}, \bibinfo{pages}{1653} (\bibinfo{year}{2023}).

\bibitem{espinosa2025slow}
\bibinfo{author}{Espinosa, M.~R.} \emph{et~al.}
\newblock \bibinfo{title}{Slow electron spin relaxation at ambient temperatures with copper coordinated by a rigid macrocyclic ligand}.
\newblock \emph{\bibinfo{journal}{Journal of the American Chemical Society}}  (\bibinfo{year}{2025}).

\bibitem{eaton2025anisotropy}
\bibinfo{author}{Eaton, S.~S.}, \bibinfo{author}{Yamabayashi, T.}, \bibinfo{author}{Horii, Y.}, \bibinfo{author}{Yamashita, M.} \& \bibinfo{author}{Eaton, G.~R.}
\newblock \bibinfo{title}{Anisotropy of spin--lattice relaxation time (t 1) for oxo-vanadium (iv) and nitrido chromium (v) porphyrins}.
\newblock \emph{\bibinfo{journal}{Journal of the American Chemical Society}}  (\bibinfo{year}{2025}).

\bibitem{lunghi2023spin}
\bibinfo{author}{Lunghi, A.}
\newblock \bibinfo{title}{Spin-phonon relaxation in magnetic molecules: theory, predictions and insights}.
\newblock In \emph{\bibinfo{booktitle}{Computational Modelling of Molecular Nanomagnets}}, \bibinfo{pages}{219--289} (\bibinfo{publisher}{Springer}, \bibinfo{year}{2023}).

\bibitem{mariano2025role}
\bibinfo{author}{Mariano, L.~A.} \emph{et~al.}
\newblock \bibinfo{title}{The role of electronic excited states in the spin-lattice relaxation of spin-1/2 molecules}.
\newblock \emph{\bibinfo{journal}{Science Advances}} \textbf{\bibinfo{volume}{11}}, \bibinfo{pages}{eadr0168} (\bibinfo{year}{2025}).

\bibitem{takahashi2011decoherence}
\bibinfo{author}{Takahashi, S.} \emph{et~al.}
\newblock \bibinfo{title}{Decoherence in crystals of quantum molecular magnets}.
\newblock \emph{\bibinfo{journal}{Nature}} \textbf{\bibinfo{volume}{476}}, \bibinfo{pages}{76--79} (\bibinfo{year}{2011}).

\bibitem{soetbeer2018dynamical}
\bibinfo{author}{Soetbeer, J.}, \bibinfo{author}{H{\"u}lsmann, M.}, \bibinfo{author}{Godt, A.}, \bibinfo{author}{Polyhach, Y.} \& \bibinfo{author}{Jeschke, G.}
\newblock \bibinfo{title}{Dynamical decoupling of nitroxides in o-terphenyl: a study of temperature, deuteration and concentration effects}.
\newblock \emph{\bibinfo{journal}{Physical Chemistry Chemical Physics}} \textbf{\bibinfo{volume}{20}}, \bibinfo{pages}{1615--1628} (\bibinfo{year}{2018}).

\bibitem{dai2021experimental}
\bibinfo{author}{Dai, Y.} \emph{et~al.}
\newblock \bibinfo{title}{Experimental protection of the spin coherence of a molecular qubit exceeding a millisecond}.
\newblock \emph{\bibinfo{journal}{Chinese Physics Letters}} \textbf{\bibinfo{volume}{38}}, \bibinfo{pages}{030303} (\bibinfo{year}{2021}).

\bibitem{pazera2023pulse}
\bibinfo{author}{Pazera, G.~J.}, \bibinfo{author}{Krzyaniak, M.~D.} \& \bibinfo{author}{Wasielewski, M.~R.}
\newblock \bibinfo{title}{Pulse sequences for manipulating the spin states of molecular radical-pair-based electron spin qubit systems for quantum information applications}.
\newblock \emph{\bibinfo{journal}{The Journal of Chemical Physics}} \textbf{\bibinfo{volume}{158}} (\bibinfo{year}{2023}).

\bibitem{shiddiq2016enhancing}
\bibinfo{author}{Shiddiq, M.} \emph{et~al.}
\newblock \bibinfo{title}{Enhancing coherence in molecular spin qubits via atomic clock transitions}.
\newblock \emph{\bibinfo{journal}{Nature}} \textbf{\bibinfo{volume}{531}}, \bibinfo{pages}{348--351} (\bibinfo{year}{2016}).

\bibitem{tlemsani2025assessing}
\bibinfo{author}{Tlemsani, I.} \emph{et~al.}
\newblock \bibinfo{title}{Assessing the robustness of the clock transition in a mononuclear s= 1 ni (ii) complex spin qubit}.
\newblock \emph{\bibinfo{journal}{Journal of the American Chemical Society}}  (\bibinfo{year}{2025}).

\bibitem{de2021exploring}
\bibinfo{author}{de~Camargo, L.~C.} \emph{et~al.}
\newblock \bibinfo{title}{Exploring the organometallic route to molecular spin qubits: The [cpti (cot)] case}.
\newblock \emph{\bibinfo{journal}{Angewandte Chemie}} \textbf{\bibinfo{volume}{133}}, \bibinfo{pages}{2620--2625} (\bibinfo{year}{2021}).

\bibitem{chicco2021controlled}
\bibinfo{author}{Chicco, S.} \emph{et~al.}
\newblock \bibinfo{title}{Controlled coherent dynamics of {[VO (TPP)]}, a prototype molecular nuclear qudit with an electronic ancilla}.
\newblock \emph{\bibinfo{journal}{Chemical Science}} \textbf{\bibinfo{volume}{12}}, \bibinfo{pages}{12046--12055} (\bibinfo{year}{2021}).

\bibitem{lunghi2019electronic}
\bibinfo{author}{Lunghi, A.} \& \bibinfo{author}{Sanvito, S.}
\newblock \bibinfo{title}{Electronic spin-spin decoherence contribution in molecular qubits by quantum unitary spin dynamics}.
\newblock \emph{\bibinfo{journal}{Journal of Magnetism and Magnetic Materials}} \textbf{\bibinfo{volume}{487}}, \bibinfo{pages}{165325} (\bibinfo{year}{2019}).

\bibitem{yang2008quantum}
\bibinfo{author}{Yang, W.} \& \bibinfo{author}{Liu, R.-B.}
\newblock \bibinfo{title}{Quantum many-body theory of qubit decoherence in a finite-size spin bath}.
\newblock \emph{\bibinfo{journal}{Physical Review B—Condensed Matter and Materials Physics}} \textbf{\bibinfo{volume}{78}}, \bibinfo{pages}{085315} (\bibinfo{year}{2008}).

\bibitem{ma2014uncovering}
\bibinfo{author}{Ma, W.-L.} \emph{et~al.}
\newblock \bibinfo{title}{Uncovering many-body correlations in nanoscale nuclear spin baths by central spin decoherence}.
\newblock \emph{\bibinfo{journal}{Nature communications}} \textbf{\bibinfo{volume}{5}}, \bibinfo{pages}{4822} (\bibinfo{year}{2014}).

\bibitem{canarie2020quantitative}
\bibinfo{author}{Canarie, E.~R.}, \bibinfo{author}{Jahn, S.~M.} \& \bibinfo{author}{Stoll, S.}
\newblock \bibinfo{title}{Quantitative structure-based prediction of electron spin decoherence in organic radicals}.
\newblock \emph{\bibinfo{journal}{The journal of physical chemistry letters}} \textbf{\bibinfo{volume}{11}}, \bibinfo{pages}{3396--3400} (\bibinfo{year}{2020}).

\bibitem{chen2020decoherence}
\bibinfo{author}{Chen, J.} \emph{et~al.}
\newblock \bibinfo{title}{Decoherence in molecular electron spin qubits: Insights from quantum many-body simulations}.
\newblock \emph{\bibinfo{journal}{The Journal of Physical Chemistry Letters}} \textbf{\bibinfo{volume}{11}}, \bibinfo{pages}{2074--2078} (\bibinfo{year}{2020}).

\bibitem{ghosh2021spin}
\bibinfo{author}{Ghosh, K.}, \bibinfo{author}{Ma, H.}, \bibinfo{author}{Onizhuk, M.}, \bibinfo{author}{Gavini, V.} \& \bibinfo{author}{Galli, G.}
\newblock \bibinfo{title}{Spin--spin interactions in defects in solids from mixed all-electron and pseudopotential first-principles calculations}.
\newblock \emph{\bibinfo{journal}{npj Computational Materials}} \textbf{\bibinfo{volume}{7}}, \bibinfo{pages}{123} (\bibinfo{year}{2021}).

\bibitem{onizhuk2021probing}
\bibinfo{author}{Onizhuk, M.} \emph{et~al.}
\newblock \bibinfo{title}{Probing the coherence of solid-state qubits at avoided crossings}.
\newblock \emph{\bibinfo{journal}{PRX Quantum}} \textbf{\bibinfo{volume}{2}}, \bibinfo{pages}{010311} (\bibinfo{year}{2021}).

\bibitem{jahn2024contribution}
\bibinfo{author}{Jahn, S.~M.}, \bibinfo{author}{Stowell, R.~K.} \& \bibinfo{author}{Stoll, S.}
\newblock \bibinfo{title}{The contribution of methyl groups to electron spin decoherence of nitroxides in glassy matrices}.
\newblock \emph{\bibinfo{journal}{The Journal of Chemical Physics}} \textbf{\bibinfo{volume}{161}} (\bibinfo{year}{2024}).

\bibitem{lenz2017quantitative}
\bibinfo{author}{Lenz, S.}, \bibinfo{author}{Bader, K.}, \bibinfo{author}{Bamberger, H.} \& \bibinfo{author}{Van~Slageren, J.}
\newblock \bibinfo{title}{Quantitative prediction of nuclear-spin-diffusion-limited coherence times of molecular quantum bits based on copper (ii)}.
\newblock \emph{\bibinfo{journal}{Chemical Communications}} \textbf{\bibinfo{volume}{53}}, \bibinfo{pages}{4477--4480} (\bibinfo{year}{2017}).

\bibitem{yamabayashi2018scaling}
\bibinfo{author}{Yamabayashi, T.} \emph{et~al.}
\newblock \bibinfo{title}{Scaling up electronic spin qubits into a three-dimensional metal--organic framework}.
\newblock \emph{\bibinfo{journal}{Journal of the American Chemical Society}} \textbf{\bibinfo{volume}{140}}, \bibinfo{pages}{12090--12101} (\bibinfo{year}{2018}).

\bibitem{onizhuk2021pycce}
\bibinfo{author}{Onizhuk, M.} \& \bibinfo{author}{Galli, G.}
\newblock \bibinfo{title}{Pycce: A python package for cluster correlation expansion simulations of spin qubit dynamics}.
\newblock \emph{\bibinfo{journal}{Advanced Theory and Simulations}} \textbf{\bibinfo{volume}{4}}, \bibinfo{pages}{2100254} (\bibinfo{year}{2021}).

\bibitem{saikin2007single}
\bibinfo{author}{Saikin, S.}, \bibinfo{author}{Yao, W.} \& \bibinfo{author}{Sham, L.}
\newblock \bibinfo{title}{Single-electron spin decoherence by nuclear spin bath: Linked-cluster expansion approach}.
\newblock \emph{\bibinfo{journal}{Physical Review B—Condensed Matter and Materials Physics}} \textbf{\bibinfo{volume}{75}}, \bibinfo{pages}{125314} (\bibinfo{year}{2007}).

\bibitem{krzystek2015high}
\bibinfo{author}{Krzystek, J.}, \bibinfo{author}{Ozarowski, A.}, \bibinfo{author}{Telser, J.} \& \bibinfo{author}{Crans, D.~C.}
\newblock \bibinfo{title}{High-frequency and-field electron paramagnetic resonance of vanadium (iv, iii, and ii) complexes}.
\newblock \emph{\bibinfo{journal}{Coordination Chemistry Reviews}} \textbf{\bibinfo{volume}{301}}, \bibinfo{pages}{123--133} (\bibinfo{year}{2015}).

\bibitem{zecevic1998dephasing}
\bibinfo{author}{Zecevic, A.}, \bibinfo{author}{Eaton, G.~R.}, \bibinfo{author}{Eaton, S.~S.} \& \bibinfo{author}{Lindgren, M.}
\newblock \bibinfo{title}{Dephasing of electron spin echoes for nitroxyl radicals in glassy solvents by non-methyl and methyl protons}.
\newblock \emph{\bibinfo{journal}{Molecular Physics}} \textbf{\bibinfo{volume}{95}}, \bibinfo{pages}{1255--1263} (\bibinfo{year}{1998}).

\bibitem{bar2012suppression}
\bibinfo{author}{Bar-Gill, N.} \emph{et~al.}
\newblock \bibinfo{title}{Suppression of spin-bath dynamics for improved coherence of multi-spin-qubit systems}.
\newblock \emph{\bibinfo{journal}{Nature communications}} \textbf{\bibinfo{volume}{3}}, \bibinfo{pages}{858} (\bibinfo{year}{2012}).

\bibitem{ye2019spin}
\bibinfo{author}{Ye, M.}, \bibinfo{author}{Seo, H.} \& \bibinfo{author}{Galli, G.}
\newblock \bibinfo{title}{Spin coherence in two-dimensional materials}.
\newblock \emph{\bibinfo{journal}{npj Computational Materials}} \textbf{\bibinfo{volume}{5}}, \bibinfo{pages}{44} (\bibinfo{year}{2019}).

\bibitem{RN178}
\bibinfo{author}{Neese, F.}, \bibinfo{author}{Wennmohs, F.}, \bibinfo{author}{Becker, U.} \& \bibinfo{author}{Riplinger, C.}
\newblock \bibinfo{title}{The orca quantum chemistry program package}.
\newblock \emph{\bibinfo{journal}{J. Chem. Phys.}} \textbf{\bibinfo{volume}{152}}, \bibinfo{pages}{Art. No. L224108} (\bibinfo{year}{2020}).

\bibitem{shim2012robust}
\bibinfo{author}{Shim, J.}, \bibinfo{author}{Niemeyer, I.}, \bibinfo{author}{Zhang, J.} \& \bibinfo{author}{Suter, D.}
\newblock \bibinfo{title}{Robust dynamical decoupling for arbitrary quantum states of a single nv center in diamond}.
\newblock \emph{\bibinfo{journal}{Europhysics Letters}} \textbf{\bibinfo{volume}{99}}, \bibinfo{pages}{40004} (\bibinfo{year}{2012}).

\bibitem{ronczka2012optimization}
\bibinfo{author}{Ronczka, M.} \& \bibinfo{author}{M{\"u}ller-Petke, M.}
\newblock \bibinfo{title}{Optimization of cpmg sequences to measure nmr transverse relaxation time t 2 in borehole applications}.
\newblock \emph{\bibinfo{journal}{Geoscientific Instrumentation, Methods and Data Systems}} \textbf{\bibinfo{volume}{1}}, \bibinfo{pages}{197--208} (\bibinfo{year}{2012}).

\bibitem{zadrozny2014multiple}
\bibinfo{author}{Zadrozny, J.~M.}, \bibinfo{author}{Niklas, J.}, \bibinfo{author}{Poluektov, O.~G.} \& \bibinfo{author}{Freedman, D.~E.}
\newblock \bibinfo{title}{Multiple quantum coherences from hyperfine transitions in a vanadium (iv) complex}.
\newblock \emph{\bibinfo{journal}{Journal of the American Chemical Society}} \textbf{\bibinfo{volume}{136}}, \bibinfo{pages}{15841--15844} (\bibinfo{year}{2014}).

\bibitem{de2020temperature}
\bibinfo{author}{de~Guillebon, T.}, \bibinfo{author}{Vindolet, B.}, \bibinfo{author}{Roch, J.-F.}, \bibinfo{author}{Jacques, V.} \& \bibinfo{author}{Rondin, L.}
\newblock \bibinfo{title}{Temperature dependence of the longitudinal spin relaxation time t 1 of single nitrogen-vacancy centers in nanodiamonds}.
\newblock \emph{\bibinfo{journal}{Physical Review B}} \textbf{\bibinfo{volume}{102}}, \bibinfo{pages}{165427} (\bibinfo{year}{2020}).

\bibitem{wood2022long}
\bibinfo{author}{Wood, B.} \emph{et~al.}
\newblock \bibinfo{title}{Long spin coherence times of nitrogen vacancy centers in milled nanodiamonds}.
\newblock \emph{\bibinfo{journal}{Physical Review B}} \textbf{\bibinfo{volume}{105}}, \bibinfo{pages}{205401} (\bibinfo{year}{2022}).

\bibitem{kuppusamy2024spin}
\bibinfo{author}{Kuppusamy, S.~K.}, \bibinfo{author}{Hunger, D.}, \bibinfo{author}{Ruben, M.}, \bibinfo{author}{Goldner, P.} \& \bibinfo{author}{Serrano, D.}
\newblock \bibinfo{title}{Spin-bearing molecules as optically addressable platforms for quantum technologies}.
\newblock \emph{\bibinfo{journal}{Nanophotonics}} \textbf{\bibinfo{volume}{13}}, \bibinfo{pages}{4357--4379} (\bibinfo{year}{2024}).

\end{thebibliography}

\end{document}